\documentclass[12pt]{article}
\begin{document}
\title{A note on anomalous Jacobians in $2+1$ dimensions}
\author{C.~D.~Fosco~\thanks{On leave from Centro At{\'o}mico
Bariloche, 8400 Bariloche, Argentina.}\\
{\normalsize\it  The Abdus Salam ICTP}\\
{\normalsize\it  Strada Costiera 11}\\ 
{\normalsize\it 34100 Trieste, Italy}}
\date{\today}
\maketitle
\begin{abstract}
\noindent  
There exist local infinitesimal redefinitions of the fermionic fields,
which may be used to modify the strength of the coupling for the
interaction term in massless $QED_3$. Under those (formally unitary)
transformations, the functional integration measure changes by an
anomalous Jacobian, which (after regularization) yields a term with
the same structure as the quadratic parity-conserving term in the
effective action.  Besides, the Dirac operator is affected by the
introduction of new terms, apart from the modification in the minimal
coupling term.  We show that the result coming from the Jacobian, plus
the effect of those new terms, add up to reproduce the exact quadratic
term in the effective action. Finally, we also write down the form a
finite decoupling transformation would have, and comment on the
unlikelihood of that transformation to yield a helpful answer to the
non-perturbative evaluation of the fermionic determinant.
\end{abstract}
\bigskip
\section{Introduction}
It has been known for quite a long time that some (non trivial) $1+1$
dimensional partition functions may be exactly evaluated by performing
a suitably chosen `decoupling' change of variables~\cite{rs}.  That
change of variables is such that, when the action is written in terms
of the new fields, there is no interaction term.  The infinitesimal
version of this transformation, on the other hand, makes it possible
to reduce the interaction term by a constant factor.  Of course, that means
that the coupled and decoupled theories are {\em classically\/}
equivalent.  Quantum mechanically, however, the need for a
regularization of the path integral measure induces (when there are
anomalies) a non trivial anomalous Jacobian~\cite{fuji}. This
Jacobian, a purely quantum artifact, takes into account effects that
would otherwise come from loop corrections in a standard perturbative
calculation, and should therefore be included into the effective action.
The infinitesimal form of the Jacobian is completely determined by the
anomaly, which in turn may be exactly calculated, and it is not
affected by radiative corrections~\cite{jackiw}. This technique may be
applied to solve some models exactly, like $QCD$ in $1+1$
dimensions~\cite{fradkin}.

Of course, the applicability of this procedure relies on the property
that there exist local changes of variables whose net effect is
tantamount to a modification of the coupling constant, something which
is far from true for an arbitrary theory. We shall deal here with the 
particular case of fermionic determinants in the presence of background 
gauge fields.  

Although it is always possible (both for the Abelian and non Abelian
cases) to construct {\em non local\/} field redefinitions, that
decouple the fields order by order in perturbation theory~\cite{fl},
local, formally unitary, transformations have not yet been used to
achieve the same goal.  Indeed, extensions of this powerful method to
higher dimensional systems are lacking, except in very particular
cases and restricted field configurations. In reference~\cite{fs},
local field redefinitions have been used to obtain the parity
violating part of the effective action in three dimensional $QED$, a
result that can also be obtained by generalized parity transformations
of the fermions~\cite{cfs}. The parity conserving part is much harder
to find by an application of this method, since it does not seem to be
related to an anomalous Jacobian.

It is our purpose in this note to show that there are, indeed,
transformations in $2+1$ dimensions that share some properties of
their $1+1$ dimensional counterparts, like locality and formal
unitarity, allowing one to derive also the quadratic parity conserving
part of the effective action. The result comes from the evaluation of
two functional determinants, both of them naively equal to one, but
nevertheless conspiring to produce the correct answer when properly
regulated.  The main difference with the $1+1$ dimensional case shall
be that the infinitesimal transformations also generate new, unwanted
interactions.  Their effect can be accounted for infinitesimally, but
we will see that, for a finite transformation, the effect is so
complicated that the method becomes not applicable. It will turn out
that the main difference between the $2+1$ and $1+1$ dimensional cases
shall be that in the former the decoupling transformation is not a
neat redefinition of the coupling, but it modifies the Dirac operator
as well. There are, however, some striking similarities, like the
fact that at least part of the effective action can be obtained from
anomalous Jacobians.

To proceed, we begin by giving a lighting review of the procedure for 
the case of a massless fermion field in the presence of an Abelian external
field, in $1+1$ dimensions. There, one wants to evaluate the Euclidean functional 
integral ${\mathcal Z}[A]$, corresponding to {\em massless\/} fermions in 
the presence of an external gauge field  $A_\mu$,
\begin{equation}\label{eq:defza}
{\mathcal Z}[A] \;=\; \int {\mathcal D}{\bar\psi} {\mathcal D}\psi \;
e^{- S[{\bar\psi},\psi;A]} \;,
\end{equation}
where 
\begin{equation}\label{eq:defs}
S[{\bar\psi},\psi;A] \;=\; \int d^2x \, {\bar\psi}\, {\mathcal D}_A \,
\psi \;,
\end{equation}
and ${\mathcal D}_A$ is the Dirac operator in the presence of an $A_\mu$ 
background field. Namely, for massless fermions in a flat spacetime,
${\mathcal D}_A\,=\, {\not\!\!D}_A$, where $D_A$ is the gauge covariant 
derivative: $D_A = \partial + i A$.

We first recall that, in any number of spacetime dimensions, we can get 
rid of the longitudinal part of the gauge field, by a proper local 
transformation of the fermions.
Indeed, decomposing $A$ into transverse and longitudinal parts, 
\begin{equation}\label{eq:adecomp}
A_\mu \;=\; A_\mu^\perp \,+\, \partial_\mu \varphi 
\end{equation}
where $\partial_\mu A^\perp_\mu = 0$, the change of fermionic variables:
\begin{equation}
\psi(x) \,\to \, \exp[-i \varphi(x) ] \psi(x) \;\;,\;\;
{\bar\psi}(x) \,=\, {\bar\psi}(x) \, \exp[ i \varphi (x)]
\end{equation}
(which is non anomalous) leads to the identity:
\begin{equation}\label{eq:ztrans}
{\mathcal Z}[A] \;=\; {\mathcal Z}[A^\perp] \;.
\end{equation}
In what follows, we shall assume that this has already been done, 
so that our starting point shall be ${\mathcal Z}[A]$ with a transverse 
$A$. We shall omit the explicit $\perp$ notation, since transversality 
of $A$ will always be assumed. 

Then, the infinitesimal change of variables
\begin{eqnarray}\label{eq:infcv}
\psi(x) &\to& \psi(x) \;+\; \xi \,\gamma_5 \alpha (x) \psi (x) \nonumber\\ 
{\bar\psi}(x) &\to& {\bar\psi}(x) \;+\; \xi \, {\bar\psi}(x) \alpha(x) 
\gamma_5 
\end{eqnarray} 
with the condition $\epsilon_{\mu\nu}\partial_\nu \alpha = A_\mu$,
changes the action $S$ as follows:
\begin{equation}\label{eq:moda}
S[{\bar\psi},\psi;A] \;\to\; S[{\bar\psi},\psi;(1-\xi)A]
\;=\; \int d^2x \; {\bar\psi} \, 
[\not\!\partial + i (1-\xi) \not \!\! A] \psi \;.
\end{equation} 
On the other hand, the measure acquires a Jacobian:
$$
J\,=\, \exp [ - \frac{\xi}{\pi} \, \int d^2x \, \alpha(x) 
\epsilon_{\mu\nu} \partial_\mu A_\nu ]
\,=\, \exp [ + \frac{\xi}{\pi} \int d^2x A_\mu A_\mu ]\;.
$$
\begin{equation}\label{eq:jac1}
\,=\, \exp [  \frac{\xi}{2\pi} \int d^2x \, 
F_{\mu\nu} \frac{1}{\partial^2} F_{\mu\nu} ]\;.
\end{equation}
When properly iterated, this procedure may be used to obtain the 
correct answer for a finite transformation~\cite{rs}. In short, there appears an extra
$\frac{1}{2}$ factor before the non local Maxwell action (we are of
course regarding $A$ as purely transverse).

In the present letter, we shall be concerned with a functional 
integral like (\ref{eq:defza}), but with $S$ now denoting the $2+1$ 
dimensional Euclidean action:
\begin{equation}\label{eq:defs3}
S[{\bar\psi},\psi;A] \;=\; \int d^3x \; {\bar\psi} {\mathcal D}_A \psi 
\;\;\;,\;\; {\mathcal D}_A = \not \!\partial + i \not \!\! A \;, 
\end{equation}
the Dirac's matrices being in a representation such that $\gamma_0 = 
\sigma_3$, $\gamma_1 = \sigma_1$, and $\gamma_2 = \sigma_2$, with $\sigma_j$ 
the standard Pauli matrices.

Regarding the actual form of the decoupling transformations, there is 
of course no equivalent to $\gamma_5$ in $2+1$ dimensions. Nevertheless,
it is evident that to modify $A_\mu$, the transformation must involve the 
$\gamma$ matrices, indeed, the same transformation we used in~\cite{fs} 
does the job. The main difference is that now we will pick up the parity 
conserving part of the corresponding Jacobian, by using a parity-conserving 
regulator. We shall see that the evaluation of that Jacobian involves 
a subtle point.  

Thus, taking into account the constraints of unitarity and locality, 
we are led to consider the infinitesimal transformations:
\begin{equation}\label{eq:ichg}
\delta \psi (x) \;=\; - \xi \, \not \! b (x) \psi (x) \;\;,\;\;
\delta {\bar\psi} (x) \;=\; - \xi {\bar\psi} \not \! b (x) \;,
\end{equation} 
where $\xi$ is infinitesimal, and $b_\mu(x)$ is a vector field.
${\mathcal Z}(A)$ cannot change under a change of variables like
(\ref{eq:ichg}), however, it yields an equivalent way of writing it:
\begin{equation}\label{eq:ch1}
{\mathcal Z}[A] \;=\; {\mathcal J}_\xi [b]  \;\times\; 
\int {\mathcal D}{\bar\psi} {\mathcal D}\psi \, 
e^{- S_\xi [{\bar\psi},\psi;A]} \;,
\end{equation}
where ${\mathcal J}_\xi$ is the Jacobian for the transformation 
(\ref{eq:ichg}), and
\begin{equation}\label{eq:ch2}
S_\xi [{\bar\psi},\psi;A] = \int d^3x {\bar\psi} \left[
\not\!\partial + i \gamma_\mu (A_\mu - \xi {\tilde f}_\mu (b))
- \xi ( b \cdot D + D \cdot b) \right] \psi
\end{equation}
with ${\tilde f}_\mu(b) = \epsilon_{\mu\nu\lambda}\partial_\nu 
b_\lambda$. It is clear that, if we choose $b_\mu$ such that
\mbox{${\tilde f}_\mu (b) \;=\; A_\mu$}, then the minimal coupling
term will be suppressed (if $\xi > 0$) by  a $(1-\xi)$ factor, although 
there have appeared two unwelcome extra terms, which were not present in the original
action.  Since $\xi$ is infinitesimal, we may also write (\ref{eq:ch1}) as:
\begin{equation}\label{eq:ch3}
{\mathcal Z}(A) \;=\; {\mathcal J}_\xi \;\;\times\;\;  {\mathcal K}_\xi 
\;\;\times\;\; {\mathcal Z}\left[ (1-\xi) A \right]\;,
\end{equation}
where
\begin{equation}\label{eq:defk}
K_\xi (A) \;=\; \det[ 1 -  \xi {\not \!\! D}^{-1} ( b \cdot D +
D \cdot b) ] 
\end{equation}
is a factor that takes care of the extra term  in the action.
Thus, to first order in $\xi$, the change in the functional integral is
determined by the two factors $J_\xi$ and $K_\xi$, of which $J_\xi$ is
a Jacobian. Note that, to first order in $\xi$, both are ill-defined,
since they involve a $0 \times \infty$.

Let us consider now each factor separately.
The Jacobian ${\mathcal J}_\xi (b)$, is
\begin{equation}\label{eq:deftr}
{\mathcal J}_\xi (b) \;=\; \left[\det( 1 - \xi \not \! b) \right]^{-2} 
\end{equation}
with a negative power because of the Grassmann character of the fields,
and a $2$ because both $\psi$ and ${\bar\psi}$ contribute by the same 
amount. 
Thus for an infinitesimal $\xi$,
\begin{equation}
{\mathcal J}_\xi \;=\; \exp[ 2 \, \xi \, {\mathcal A}  ]
\end{equation}
with
\begin{equation}\label{eq:trace}
{\mathcal A} \;=\; {\rm Tr}[ \not \! b] \;.
\end{equation}
The (functional and Dirac) trace of $\not \! b$ suffers from the same kind
of ill behaviour the trace of $\gamma_5$ times a scalar field does in 
$1+1$ dimensions. Hence the need for a regularization. However, a 
delicate question arises here, since a naive calculation easily
misleads one to  a wrong answer: let ${\mathcal A}_\Lambda$, the 
regularized version of (\ref{eq:trace}) be
\begin{equation}\label{eq:rtrace}
{\mathcal A}_\Lambda \;\equiv\; {\rm Tr} \left[\not \! b \, 
f(-\frac{{\not\!\! D}^2}{\Lambda^2}) \right] 
\end{equation}
where $\Lambda$ is an UV cutoff, and $f$ a regularizing function, verifying:
\begin{equation}
f(0)= 1 \;, \;\;\; f(\pm \infty) = f'(\pm \infty) 
= \ldots f^{(k)} (\pm \infty) = \ldots = 0 \;.
\end{equation} 
Taking, for example, $f(x) = \frac{1}{1 + x}$, we see after some standard
manipulations that the only term that does not vanish when 
$\Lambda \to \infty$ is given (in Fourier space) by:
\begin{equation}
{\mathcal A}_\Lambda \;=\; i \int \frac{d^3 k}{(2\pi)^3} \;
{\tilde b}_\mu (-k) \; I(k)\, \epsilon_{\mu\nu\lambda} k_\nu 
\; {\tilde A}_\lambda (k)  
\end{equation}
where the tilde denotes the Fourier transformed of the corresponding
object, while $I(k)$ is defined by the expression:
\begin{equation}
I(k)\,= - 2 \Lambda^2 \, \int \frac{d^3p}{(2\pi)^3} 
\frac{1}{(p^2+\Lambda^2) [(p + k)^2 + \Lambda^2]} \;.
\end{equation}
Being the momentum integral convergent, we evaluate it and consider its
$\Lambda \to \infty$ limit. This leads to a result diverging linearly
with $\Lambda$. Using afterwards the relation between $A_\mu$ 
and $b_\mu$, the upshot of this procedure is a {\em divergent\/} 
answer for ${\mathcal A}$:
\begin{equation}\label{eq:ntrace}
{\mathcal A}_\Lambda \;\sim\; \Lambda \int d^3x \, A_\mu A_\mu
\;\;,\;\;\; \Lambda \sim \infty \;,
\end{equation}
which is dimensionally correct: in our conventions, the mass dimension 
of $A$ is $1$.  Note that no perturbative approximation in $A$ has been 
used, the $\Lambda \to \infty$ limit selects that contribution, and higher 
order terms simply fade away, as negative powers of $\Lambda$.

The reason for the (wrong) result (\ref{eq:ntrace}), is that it 
is obtained by regarding  $b_\mu$ as a local field, while it is related
non locally to $A_\mu$.  Indeed, the relation between $A$ and $b$ is 
{\em momentum dependent\/}, since 
\begin{equation}
b_\mu \;=\; - \frac{1}{\partial^2} \epsilon_{\mu\nu\lambda} 
\partial_\nu A_\lambda \;,
\end{equation}
and this relation must be used {\em before\/} doing the momentum 
integration, since we really want to know the Jacobian as a function of 
$A_\mu$, which is a true local field~\footnote{Note that the locality
of the fermionic field transformation is beyond any question:
the new field value at point $x$ does not involve the values of the fermionic 
field at different points.}. 
Using this relation  before the integration, there appears a quite 
different momentum space integral, since now we may write 
${\mathcal A}_\Lambda$ as follows:
\begin{equation}
{\mathcal A}_\Lambda \;=\; \int \frac{d^3 k}{(2\pi)^3} \,
{\tilde F}_{\mu\nu} (-k) \, L(k) \, {\tilde F}_{\mu\nu} (k) 
\end{equation}
where ${\tilde F}_{\mu\nu}(k) = i 
(k_\mu {\tilde A}_\nu - k_\nu {\tilde A}_\mu)$,  $L(k)$ is 
defined by:
\begin{equation}
L(k)\,= - \Lambda^2 \, \int \frac{d^3p}{(2\pi)^3} 
\frac{1}{p^2 (p^2+\Lambda^2) [(p + k)^2 + \Lambda^2]} \;,
\end{equation}
and we have only kept terms that do not vanish when $\Lambda \to
\infty$.

Surprisingly enough, $L$ is {\em convergent\/} when $\Lambda \to \infty$:
\begin{equation}\label{eq:sint}
L(k) \;\to \; - \frac{1}{|k|}\;\;,\;\; \Lambda \to \infty\;.
\end{equation}
In coordinate space, the answer for ${\mathcal J}_\xi$ is then:
\begin{equation}
{\mathcal J}_\xi \;=\; \exp \left[- \frac{\xi}{4} \, \int d^2 \, 
F_{\mu\nu}(A) \frac{1}{\sqrt{-\partial^2}} F_{\mu\nu}(A) \right] \;,
\end{equation}
which, except for the factor, has the same structure of the exact
 quadratic parity conserving term, known to be: 
\begin{equation}\label{eq:exact}
( {\mathcal Z}[A] )_{quad.}\;=\; \exp 
\left[- \frac{1}{4} \int d^3x \, F_{\mu\nu} 
\frac{1}{16\sqrt{-\partial^2}}  F_{\mu\nu}  \right] \;.
\end{equation}
The ${\mathcal K}_\xi$ factor may be evaluated in a  very similar way, 
namely,  by using the regulating function $f(x) = \frac{1}{1+x}$ and keeping
only the non-vanishing terms when $\Lambda \to \infty$.
The result has of course the same structure, and differs from the contribution
of ${\mathcal J}_\xi$ only by a factor:
\begin{equation}
{\mathcal K}_\xi \;=\; \exp \left[ \frac{\xi}{4} \int d^3x \,
F_{\mu\nu} \frac{7}{8\sqrt{-\partial^2}}  F_{\mu\nu} \right]\;
\end{equation}
Putting together both ${\mathcal J}_\xi$ and ${\mathcal K}_\xi$, we see that:
\begin{equation}
\ln {\mathcal Z}[A] - \ln {\mathcal Z}[(1-\xi) A] \;=\;
\int d^3 x \left[-
\frac{\xi }{4} \int d^3x \, F_{\mu\nu} \frac{1}{8\sqrt{-\partial^2}}
F_{\mu\nu} \,+\, {\mathcal O}(\xi^2) \right]\;,
\end{equation}
which is the exact result we where looking for, since it leads to 
(\ref{eq:exact}) by a simple integration. 

We end up by considering  the form a finite version of the decoupling
transformation might have. We apply it to the free Dirac operator,
$\not \!\! \partial$, to see whether the interacting operator may be reproduced
or not by a change of variables, as it happens in $1+1$ dimensions. 
We expect `spurious' terms, of course, but it is interesting to see their
explicit form. They will be, in fact, the real obstruction for a local,
finite decoupling transformation to exist.

To be unitary, and to reduce to the infinitesimal transformations we have 
considered, we must have:
\begin{eqnarray}\label{eq:finite}
\psi (x) & \to & [ a(x) \,- \, \not\!b(x) ] \psi (x) \nonumber\\
{\bar\psi}(x) &\to &  {\bar\psi}(x) [ a(x) \,- \, \not\!b(x) ] 
\end{eqnarray}
where $a^2 - b_\mu b^\mu  = 1$. Under this transformation,  the 
free action written in the new variables may be written in the following 
form:
$$
S \;=\; \int d^3 x {\bar\psi}(x) \left\{ \not \! \partial \,+\,
[ \partial \cdot b b_\mu - i a {\tilde f}_\mu + b \cdot (\partial b_\mu)
 - i \epsilon_{\mu\nu\lambda} b_\nu \partial_\lambda a ] \gamma_\mu  
\right.
$$
\begin{equation}
\left. +\, [ i \epsilon_{\mu\nu\lambda} b_\mu \partial_\nu b_\lambda 
- (a \partial \cdot b + (b \cdot \partial a)] \right\}\;.
\end{equation}
One can immediately see that it is not possible to fulfill at the same time
the conditions required to have the proper coupling term and no extra
term at the same time. The conditions to have the coupling term with 
an external field $A_\mu$ cannot me met, since the corresponding equations are not
compatible. This negative result is to be expected, since we knew that
the infinitesimal version already produces new terms in the action.

We may then conclude by saying that the main obstruction to the application
of the local decoupling transformations in higher dimensions is 
`kinematical', since there are no (local) transformations which amount to just 
a redefinition of the coupling. The fact that part of the effective action may
indeed be obtained from an anomalous change of variables is, surprisingly, 
true.

\section*{Acknowledgements:}
This work was supported by a Fundaci{\'o}n Antorchas grant, and by an ICTP 
Regular Associateship scheme. 
The author acknowledges Prof.\ F.\ A.\ Schaposnik for reading the manuscript.

\end{document}